\documentclass[letter,twocolumn]{jpsj2}
\title{Extremely Large and Anisotropic Upper Critical Field and the Ferromagnetic Instability in UCoGe}

\author{%
Dai~\textsc{Aoki}$^1$\thanks{E-mail address: aokidai@gmail.com}, %
Tatsuma~D.~\textsc{Matsuda}$^{1,2}$
Valentin~\textsc{Taufour}$^1$,
Elena~\textsc{Hassinger}$^1$,
Georg~\textsc{Knebel}$^1$, and
Jacques~\textsc{Flouquet}$^1$
}

\inst{%
$^1$INAC/SPSMS, CEA-Grenoble, 17 rue des Martyrs, 38054 Grenoble, France\\
$^2$Advanced Science Research Center, Japan Atomic Energy Agency, Tokai, Ibaraki 319-1195, Japan\\
}

\abst{%
Magnetoresistivity measurements with fine tuning of the field direction on high quality single crystals of the 
ferromagnetic superconductor UCoGe show anomalous anisotropy of the upper critical field $H_{\rm c2}$.
$H_{\rm c2}$ for $H\parallel b\mbox{-axis}$ ($H_{\rm c2}^b$) in the orthorhombic crystal structure
is strongly enhanced with decreasing temperature with an S-shape and reaches nearly $20\,{\rm T}$ at $0\,{\rm K}$.
The temperature dependence of $H_{\rm c2}^a$ shows upward curvature with a low temperature value exceeding $30\,{\rm T}$,
while $H_{\rm c2}^c$ at $0\,{\rm K}$ is very small ($\sim 0.6\,{\rm T}$).
Contrary to conventional ferromagnets, 
the decrease of the Curie temperature with increasing field for $H\parallel b\mbox{-axis}$
marked by an enhancement of the effective mass of the conduction electrons 
appears to be the origin of the S-shaped $H_{\rm c2}^b$ curve.
These results indicate that the field-induced ferromagnetic instability or magnetic quantum criticality reinforces superconductivity.
}

\kword{unconventional superconductivity, spin-triplet state, upper critical field, quantum criticality, effective mass, ferromagnetism, UCoGe}

\begin{document}
\maketitle
The coexistence of superconductivity (SC) and ferromagnetism (FM) has attracted much attention,
since the exotic SC state based on spin-triplet pairing mediated by longitudinal spin fluctuations 
is expected.~\cite{Fay80} 
The first example was discovered in UGe$_2$ under pressure, 
where $T_{\rm sc}$ is much lower than $T_{\rm Curie}$.~\cite{Sax00}
The SC phase exists only in the FM phase, and SC disappears in the paramagnetic (PM) phase above the critical pressure $P_{\rm c}$.
Soon after that, SC was found at ambient pressure in the weak ferromagnet URhGe.~\cite{Aok01}
$T_{\rm sc}$ ($=0.25\,{\rm K}$) is much lower than $T_{\rm Curie}$ ($=9.5\,{\rm K}$).
The upper critical field $H_{\rm c2}$ exceeds the Pauli paramagnetic limit, 
thus it is believed that the spin-triplet state with equal-spin pairing is realized.
Recently, reentrant superconductivity (RSC) was found between $8$ and $13\,{\rm T}$ 
when the field is applied 
along the $b$-axis of the orthorhombic TiNiSi-type structure (space group: \textit{Pnma}).~\cite{Lev05}
With increasing field, the magnetic moment gradually tilts from the $c$-axis (easy-magnetization axis) to $b$-axis.
Finally the moment is suddenly aligned to the $b$-axis at the spin reorientation field $H_{\rm R}$.
The field-induced critical magnetic fluctuations induce RSC.
Recently, we have observed the enhancement of effective mass in the RSC phase, 
and explained the emergence of RSC.~\cite{Miy08,Miy09}

A newcomer, UCoGe, which crystallizes in the same structure as URhGe, was recently reported.~\cite{Huy07}
UCoGe is a weak ferromagnet with $T_{\rm Curie}\sim 3\,{\rm K}$ and the ordered moment $\mu_0 = 0.07\,\mu_{\rm B}/{\rm U}$.
$T_{\rm sc}$ ($\sim 0.6\,{\rm K}$) is larger than that in URhGe.
Since $T_{\rm Curie}$ is low, one can naively consider that UCoGe is close to the quantum critical point.
Indeed, our previous measurement shows that $T_{\rm Curie}$ is 
immediately suppressed by applying a small pressure ($P_{\rm c}\sim 1\,{\rm GPa}$).~\cite{Has08_UCoGe}
Contrary to the case of UGe$_2$ where SC exists only in the FM domain, 
SC survives even in the PM phase with the maximum $T_{\rm sc}$ ($\simeq 0.75\,{\rm K}$) around $P_{\rm c}$.
A new theory from symmetry considerations was developed in order to explain 
the temperature-pressure phase diagram.~\cite{Min08}
$H_{\rm c2}$ at ambient pressure shows strong anisotropy. 
$H_{\rm c2}$ for $H\parallel a$ ($H_{\rm c2}^a$) and $b\mbox{-axis}$ ($H_{\rm c2}^b$)
reveal almost the same temperature dependence, 
exceeding the Pauli paramagnetic limit.~\cite{Huy08}
Because the characteristics of UCoGe are similar to those of URhGe,
RSC or field-tuned ferromagnetic instability is expected.
There are, however, no reports on this up to now.

Here we report the results of resistivity measurements with high quality single crystals of UCoGe
under magnetic fields with fine tuning of the field direction.
Contrary to the previous report, for the perfect field alignment along $b$-axis,
$H_{\rm c2}^b$ is strongly enhanced on cooling at $T/T_{\rm sc}=0.65$, 
revealing an S-shaped behavior.
$H_{\rm c2}^a$ is similar to $H_{\rm c2}^b$ down to $T/T_{\rm sc}=0.65$,
but continuously increases with upward curvature with decreasing temperature.
$T_{\rm Curie}$ for $H\parallel b\mbox{-axis}$ is connected to the S-shaped $H_{\rm c2}$ phase diagram,
where the effective mass is strongly enhanced, indicating the existence of a field-tuned ferromagnetic instability around $14\,{\rm T}$.
Compared to the results of URhGe where there is a clear separation between low field SC and RSC,
here we believe the S-shaped $H_{\rm c2}^b$ in UCoGe is due to the mixing of RSC phase and low field SC phase.

High quality single crystals were grown using the Czhochralski method in the tetra-arc furnace.
The details will be published elsewhere.

The electrical resistivity was measured employing the four probe AC method in a dilution refrigerator 
at high fields up to $16\,{\rm T}$ with a sample rotation mechanism.
Two single crystals were used for the present study. 
Sample\#1 is for the current along $a$-axis and the field direction from $b$ to $a$-axis. 
Sample\#2 is for the current along $c$-axis and the field direction from $a$ to $c$-axis.
The electrical currents were $50\,\mu{\rm A}$ for both samples.
The field orientation was checked with a Hall sensor in order to control the precise field angle.
The sample quality was tested by resistivity and specific heat measurements.
The resistivity shows a clear kink due to the ferromagnetic transition at $T_{\rm Curie}=2.6\,{\rm K}$,
indicating the high quality of the present sample.
The residual resistivity ratio for both samples is 30, 
which is comparable to the previous report on a single crystal sample.~\cite{Huy08}
The specific heat shows two clear jumps at $T_{\rm Curie}=2.6\,{\rm K}$ and $T_{\rm sc}=0.44\,{\rm K}$,
indicating the high quality sample and bulk superconductivity, which will be published elsewhere. 

\begin{figure}[htb]
\begin{center}
\includegraphics[width=0.7 \hsize,clip]{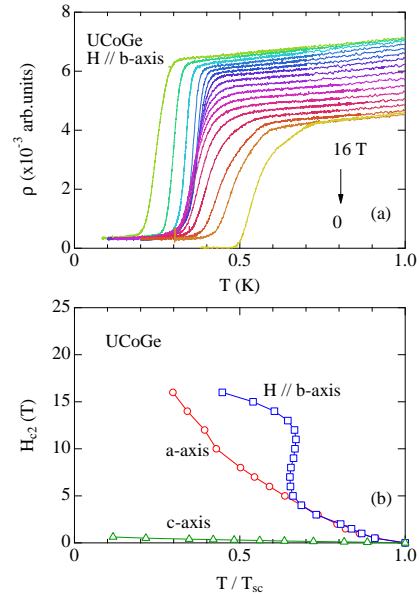}
\end{center}
\caption{(Color online) (a)Temperature dependence of the electrical resistivity at various fields from $16$ to $0\,{\rm T}$ with $1\,{\rm T}$ steps for $H \parallel \mbox{b-axis}$ in sample\#1. 
(b)Temperature dependence of the upper critical fields for $H \parallel {\rm a}$, b and c-axis. The temperature is normalized by the superconducting critical temperature $T_{\rm sc}$ at zero field.}
\label{fig:UCoGe_rho_Hc2}
\end{figure}

Figure~\ref{fig:UCoGe_rho_Hc2}(a) shows the low temperature resistivity of sample\#1 under the various fields for $H\parallel b\mbox{-axis}$
The field direction is precisely adjusted by rotating the sample.
At zero field, the resistivity starts to drop due to the superconductivity at onset temperature of $0.72\,{\rm K}$, and becomes zero at $0.48\,{\rm K}$.
The temperature corresponding to the half drop of resistivity is $0.55\,{\rm K}$, which will be defined as $T_{\rm sc}$ hereafter.
$T_{\rm sc}$ decreases with increasing fields.
Surprisingly, the resistivity drop starts to become sharp above $5\,{\rm T}$,
and $T_{\rm sc}$ at $11\,{\rm T}$ is slightly larger than that at $6\,{\rm T}$.
Above $11\,{\rm T}$, $T_{\rm sc}$ again decreases with fields,
however SC survives even at the maximum field of $16\,{\rm T}$.
Above $1\,{\rm T}$ there is a small remanent resistivity below $T_{\rm sc}$.
This is most likely due to the sample which contains a small amount of $c$-axis components against the field direction, 
since the resistivity shows double steps below $0.6\,{\rm T}$,
and $H_{\rm c2}$ corresponding to the remanent resistivity drop to zero
is approximately the same as that for $H \parallel c\mbox{-axis}$.
Nevertheless, the overall behavior of $H_{\rm c2}$ for $H\parallel b\mbox{-axis}$ 
is not affected by these small components aligned to $c$-axis.

The phase diagram of $H_{\rm c2}^b$ is shown in Fig.~\ref{fig:UCoGe_rho_Hc2}(b), 
together with those of $H_{\rm c2}^a$ and $H_{\rm c2}^c$.
Since the data for $H \parallel a$ and $c$-axis were obtained by using sample{\#2} with slightly higher $T_{\rm sc}=0.71\,{\rm K}$,
the temperature in Fig.~\ref{fig:UCoGe_rho_Hc2}(b) is normalized by $T_{\rm sc}$ at zero field for clarity.
At $T/T_{\rm sc}\simeq 0.65$, $H_{\rm c2}^b$ is strongly enhanced and shows an S-shaped temperature dependence.
On the other hand, $H_{\rm c2}^a$ shows no abrupt enhancement,
but increases with upward curvature on cooling.
It is noted that 
$H_{\rm c2}^a$ is almost the same as $H_{\rm c2}^b$ down to $T/T_{\rm sc}=0.65$.
Contrary to these high $H_{\rm c2}^a$ and $H_{\rm c2}^b$, 
$H_{\rm c2}^c$ is much smaller.
$H_{\rm c2}^c$ curve reveals the slight upward curvature and $H_{\rm c2}^c (0)$ at $0\,{\rm K}$ is $0.6\,{\rm T}$.

Assuming the weak-coupling BCS model with $g$-factor ($g=2$), the Pauli paramagnetic limiting field is $H_{\rm P}=1$--$1.3\,{\rm T}$,
which is much lower than $H_{\rm c2}^a(0)$ and $H_{\rm c2}^b(0)$,
indicating that $H_{\rm c2}^a(0)$ and $H_{\rm c2}^b(0)$ are governed by the orbital limit without Pauli paramagnetic effect.
The pairing symmetry with $H_{\rm c2}$ exceeding Pauli limiting field was theoretically 
discussed in the ferromagnetic superconductor UGe$_2$.~\cite{Mac01}
According to this theory, the non-unitary spin-triplet state must be realized.
In the same manner, the present experimental results strongly support that UCoGe is an unconventional superconductor 
with the non-unitary spin-triplet pairing symmetry.
The large anisotropy of $H_{\rm c2}$ is most likely ascribed to the anisotropy of the field-dependent effective mass of the conduction electrons, which will be discussed later.

\begin{figure}[htb]
\begin{center}
\includegraphics[width=0.7 \hsize,clip]{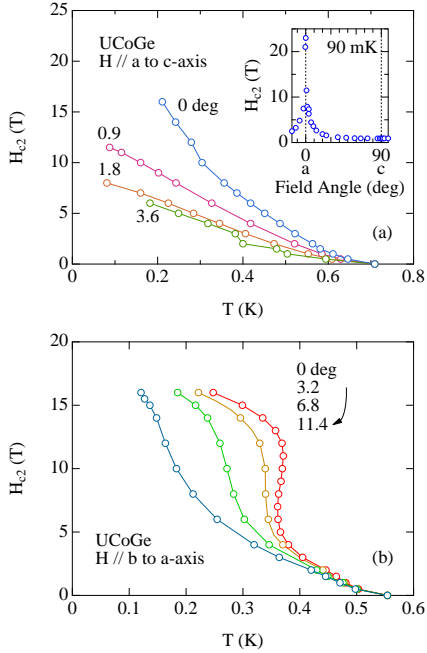}
\end{center}
\caption{(Color online) (a)Temperature dependence of the upper critical field $H_{\rm c2}$ close to $H\parallel \mbox{a-axis}$ in sample\#2. The field direction is tilted from the $a$ to $c$-axis. The inset shows the angular dependence of $H_{\rm c2}$ at $90\,{\rm mK}$. The values of $H_{\rm c2}$ greater than $16\,{\rm T}$ are the results of linear extrapolations to $90\,{\rm mK}$.
(b)Temperature dependence of $H_{\rm c2}$ close to $H\parallel b\mbox{-axis}$ in sample\#1. The field direction is tilted from the $a$ to $c$-axis.}
\label{fig:UCoGe_Hc2_a_b}
\end{figure}

The present $H_{\rm c2}$ phase diagram at least for $H \parallel a$ and $b\mbox{-axis}$
is very different to the previous results,~\cite{Huy08}
in which $H_{\rm c2}^a$ and $H_{\rm c2}^b$ show almost the same temperature variations with slight upward curvatures.
Previously reported values for 
$H_{\rm c2}^a(0)$ and $H_{\rm c2}^b(0)$ ($\approx 5\,{\rm T}$) are smaller than the present results,
while the present result of $H_{\rm c2}^c$ is in good agreement with the previous result.
This is most likely due to the fine tuning of field orientation and the sample quality.
Figure~\ref{fig:UCoGe_Hc2_a_b}(a) shows the $H_{\rm c2}$ curves close to $H \parallel a$.
The field angles from $a$ to $c$-axis are denoted.
In the case that the field direction is optimized to $H\parallel a\mbox{-axis}$,
$H_{\rm c2}^a$ curve shows the upward curvature, and
$H_{\rm c2}^a$ reaches more than $20\,{\rm T}$ at $90\,{\rm mK}$ from the linear extrapolation between $12$ and $16\,{\rm T}$.
At $0\,{\rm K}$, it would be close to $H_{\rm c2}^a(0) \sim 30\,{\rm T}$.
This extraordinarily high $H_{\rm c2}^a(0)$ is strongly suppressed
when the field direction is slightly tilted from the $a$ to $c$-axis.
$H_{\rm c2}$ for $0.9\,{\rm deg}$, for example, is $11.5\,{\rm T}$ at $90\,{\rm mK}$ and $13.5\,{\rm T}$ at $0\,{\rm K}$.
As the field direction is tilted further, $H_{\rm c2}(0)$ decreases further and the upward curvature becomes milder.
The inset of Fig.\ref{fig:UCoGe_Hc2_a_b}(a) shows the angular dependence of $H_{\rm c2}$ at $90\,{\rm mK}$
from $H\parallel a$ to $c$-axis.
$H_{\rm c2}$ shows a sharp peak at the field direction aligned perfectly to $a$-axis.
Despite the fact that there is no Pauli paramagnetic effect,
the angular dependence of $H_{\rm c2}$ at $90\,{\rm mK}$ cannot be fitted
by the conventional effective-mass model based on the anisotropic Fermi surface.
This means that there should be other origins which govern the anisotropy of $H_{\rm c2}$ at low temperature.
The main new event is the proximity of UCoGe to the ferromagnetic instability 
with obviously a strong feedback from the direction of the applied magnetic field.

Figure~\ref{fig:UCoGe_Hc2_a_b}(b) shows the temperature dependence of $H_{\rm c2}$ 
in the situation where the field is tilted from the $b$ to $a$-axis.
Contrary to the results from $H\parallel a$ to $c$-axis, SC is more robust.
The S-shaped $H_{\rm c2}$ curve is smeared out by increasing the field angle,
but the main tendency remains even at $11.4\,{\rm deg}$.
Furthermore, $H_{\rm c2}(0)$ seems to increases with the field angle,
that is, $H_{\rm c2}(0)$ at $H\parallel b\mbox{-axis}$ is close to $20\,{\rm T}$,
while $H_{\rm c2}(0)$ at $11.4\,{\rm deg}$ would be close to $25\,{\rm T}$.
Since the $H_{\rm c2}(0)$ for $H\parallel a\mbox{-axis}$ is very large ($\sim 30\,{\rm T}$) with the upward curvature,
we speculate that the S-shaped $H_{\rm c2}$ curve for $H\parallel b\mbox{-axis}$ gradually disappears 
and $H_{\rm c2}(0)$ increases with increasing field angle, 
and then the huge $H_{\rm c2}^a(0)$ is finally obtained.

\begin{figure}[htb]
\begin{center}
\includegraphics[width=0.7 \hsize,clip]{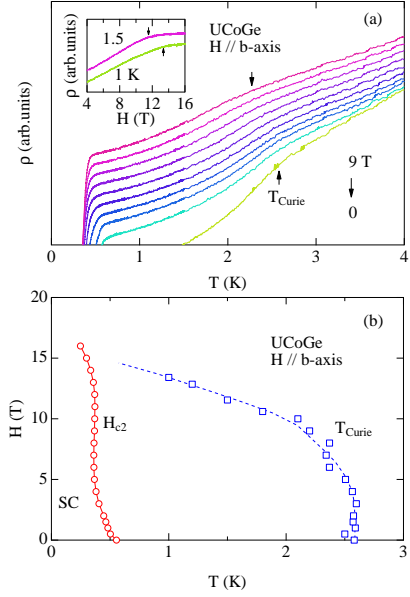}
\end{center}
\caption{(Color online) (a)Temperature dependence of the resistivity for $H \parallel \mbox{b-axis}$ at various fields from $9$ to $0\,{\rm T}$ every $1\,{\rm T}$ step. 
The data are vertically shifted for clarity. 
The inset shows the field dependence of the resistivity for $H\parallel \mbox{b-axis}$ at constant temperatures of $1.5$ and $1\,{\rm K}$
(b)Field--Temperature phase diagram for $H\parallel \mbox{b-axis}$.}
\label{fig:UCoGe_rho_phase_high}
\end{figure}

Let us look to the high temperature part of the resistivity in the normal state.
Figure~\ref{fig:UCoGe_rho_phase_high}(a) shows the temperature dependence of resistivity for $H\parallel b\mbox{-axis}$ at various fields.
$T_{\rm Curie}$ at zero field is clearly detected as a bend of resistivity at $2.6\,{\rm K}$, 
as shown by an arrow.
With increasing fields, the bend corresponding to $T_{\rm Curie}$ shifts to lower temperature.
This is quite different from the field behavior of conventional ferromagnets,
in which $T_{\rm Curie}$ seems to increase with field but the FM anomaly is rapidly smeared out;
it cannot be defined under high magnetic fields, 
because the FM transition at zero field becomes the crossover from PM state to FM state.
As shown in the inset of Fig.~\ref{fig:UCoGe_rho_phase_high}(a),
the field dependence of the resistivity at constant temperatures shows a broad bend at high field, 
which shifts to lower field with increasing temperature.
The field-temperature phase diagram for $H\parallel b\mbox{-axis}$ is shown in Fig.~\ref{fig:UCoGe_rho_phase_high}(b).
Here we refer to the critical temperature as $T_{\rm Curie}$, 
since its extrapolation to zero field corresponds to the establishment of the FM order.
$T_{\rm Curie}$ or the field corresponding to $T_{\rm Curie}$ 
is followed by the 2nd derivative of the resistivity against temperature or field.
$T_{\rm Curie}$ is initially invariant with increasing field and shifts to lower temperature at high fields.
Interestingly, $T_{\rm Curie}$ extrapolated to $0\,{\rm K}$ seems to be connected to the collapse of SC phase.
It should be noted that when the field direction is slightly tilted from $b$-axis,
the anomaly due to $T_{\rm Curie}$ is rapidly smeared out.

\begin{figure}[htb]
\begin{center}
\includegraphics[width=0.7 \hsize,clip]{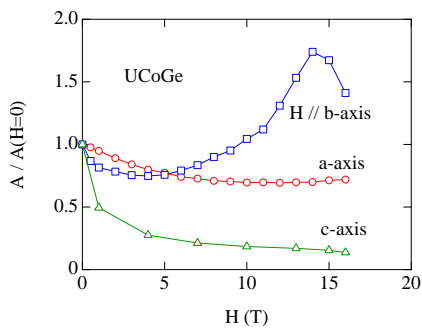}
\end{center}
\caption{(Color online) Field dependence of the normalized $A$ coefficient of resistivity for $H\parallel \mbox{a}$, b and c-axis.}
\label{fig:UCoGe_A_coef}
\end{figure}

In order to clarify the origin of the unusual $H_{\rm c2}$ curves, we have analyzed the resistivity data in the normal state
and obtained the field dependence of the $A$ coefficient normalized by the $A$ coefficient at zero field, as shown in Fig~\ref{fig:UCoGe_A_coef}.
The resistivity at constant fields follows the quadratic temperature dependence, namely $\rho = \rho_0 + AT^2$, 
in the whole field range at least below $1\,{\rm K}$.
At first glance, $A$ for $H \parallel c\mbox{-axis}$ is strongly suppressed with increasing field,
On the other hand, $A$ for $H\parallel a\mbox{-axis}$ slightly decreases but remains large.
$A$ for $H\parallel b\mbox{-axis}$ initially decreases, but then increases and becomes maximum around $14\,{\rm T}$,
and finally decreases.
If we assume the validity of the Kadowaki-Woods relation ($A \propto \gamma^2$), where $\gamma$ is the electronic specific heat coefficient, Figure~\ref{fig:UCoGe_A_coef} corresponds to the
field dependence of the square value of the effective mass, namely $A \propto m^\ast{}^2$.

According to the McMillan-type formula~\cite{Mcm68}, 
which is also applicable to spin-triplet superconductivity mediated by spin fluctuations,\cite{Fay80,Kir01} 
$T_{\rm sc}$ is determined by the following equation,
$T_{\rm sc}\propto \exp[-(\lambda + 1)/\lambda]$, 
where $\lambda$ is the coupling constant related with $m^\ast$ and the band mass $m_{\rm B}$ 
via the expression $m^\ast = (1+\lambda) m_{\rm B}$.
Thus $T_{\rm sc}$ will increase, if the correlation of the conduction electrons is enhanced.
In addition, $H_{\rm c2}$ in the spin-triplet superconductor is determined by the orbital limit,
thus, $H_{\rm c2}$ is expressed as $H_{\rm c2}=\psi_0/2\pi\xi_0{}^2$, 
where $\psi_0$ and $\xi_0$ are the quantum fluxoid and the coherence length, respectively.
Since $\xi_0$ is related to the Fermi velocity $v_{\rm F}^{}$ and $T_{\rm sc}$
as $\xi_0 \sim \hbar v_{\rm F}^{}/k_{\rm B}T_{\rm sc}$,
$H_{\rm c2}$ is simply described as $H_{\rm c2} \sim (m^\ast T_{\rm sc})^2$.
Therefore, if $m^\ast$ is enhanced as a function of field, 
$H_{\rm c2}$ is also enhanced as well as $T_{\rm sc}$ and
the SC state can be stabilized even at high fields.
Here it is noted that we assume the Fermi surface and $m_{\rm B}$ are invariant against the field, 
and the enhancement of $T_{\rm sc}$ depends on the values of $m^\ast$.
In addition, $T_{\rm sc}$ here is the hypothetical value at zero field.

Taking into account these points, 
the strong suppression of $A$ at low field for $H\parallel c\mbox{-axis}$ 
agrees with the low value of $H_{\rm c2}$.
For $H\parallel a\mbox{-axis}$, $A$ remains high with a slight decrease, 
and thus the calculated $H_{\rm c2}$ keeps a large value.
If the applied field is lower than the calculated $H_{\rm c2}$, the SC state survives.
The large enhancement of $A$ for $H\parallel b\mbox{-axis}$ must be associated with a large increase of $T_{\rm sc}$ and $H_{\rm c2}$,
and thus explains the S-shaped $H_{\rm c2}^b$ curve.
After $A$ becomes maximum at $14\,{\rm T}$, $A$ is strongly suppressed with increasing field.
Thus the SC state will be predicted to collapse immediately, which is indeed observed in $H_{\rm c2}^b$ curve.

Let us compare the present results in UCoGe with those in URhGe.
First of all, the phase diagram for $H\parallel b\mbox{-axis}$ in UCoGe is similar to that in URhGe.
The phase diagram in URhGe shows that
$T_{\rm Curie}$ ($=9.5\,{\rm K}$) is initially constant with increasing field and then gradually decreases.
Finally $T_{\rm Curie}$ is connected to the $H_{\rm R}$ ($\simeq 12\,{\rm T}$) at $0\,{\rm K}$,
at which the magnetic moment reorients to the $b$-axis.
In UCoGe, a similar phase diagram, as shown in Fig.~\ref{fig:UCoGe_rho_phase_high}(b), was obtained,
although the spin reorientation in UCoGe is not confirmed yet.
In addition, the magnetoresistance of URhGe shows a sharp peak at $H_{\rm R}$, while in UCoGe,
a broad bend was observed as shown in the inset of Fig.~\ref{fig:UCoGe_rho_phase_high}(a).
This difference may be ascribed to the small moment in UCoGe. 
Other possible reasons are the sample quality or small mis-alignment against the field.

In UCoGe, it is natural to consider that the S-shaped $H_{\rm c2}$ curve is closely related to the reduced $T_{\rm Curie}$.
$T_{\rm sc}$ is much larger than that in URhGe, because UCoGe is close to the FM instability.
Therefore, the low field SC phase is enlarged in UCoGe, compared to that in URhGe.
Then, the low field SC phase is combined with the RSC.
In reality, the S-shaped SC phase is realized in UCoGe, as a consequence of the combination of two superconducting domes.

Furthermore, the extremely high $H_{\rm c2}$ for $H\parallel a\mbox{-axis}$ in UCoGe is also explained 
from the comparison to URhGe.
It is reported that the RSC dome in URhGe shifts to higher field 
by tilting the field angle from $b$ to $a\mbox{-axis}$, linked to $H_{\rm R}$
which increases with $1/\cos\theta$-dependence ($\theta$: field angle from $b$ to $a$-axis).~\cite{Lev07}
That is, the RSC dome shows a diverging behavior towards extremely high field by tilting the field angle to $a$-axis.
In UCoGe, we obtained a very high $H_{\rm c2}$ value ($\simeq 30\,{\rm T}$) for $H\parallel a\mbox{-axis}$.
As the SC phase of UCoGe is generated from the combination of low field SC phase and RSC,
the present large $H_{\rm c2}^a$ corresponds to the divergent behavior of RSC in URhGe.
Of course, the small mis-orientation to the $c$-axis makes a strong drop of $H_{\rm c2}$,
hence, it is difficult to obtain the real divergent behavior experimentally.

The pressure-induced ferromagnetic superconductor UGe$_2$ also shows the S-shaped $H_{\rm c2}$ curve
at $1.35\,{\rm GPa}$ when the field is applied along the easy-magnetization axis ($a$-axis).~\cite{She01}
The origin is, however, different from those for UCoGe and URhGe.
At $P=1.35\,{\rm GPa}$ ($P_{\rm x} < P < P_{\rm c}$), 
The FM state in UGe$_2$ changes from FM1 (weakly polarized phase) to FM2 (strongly polarized phase) with increasing field.
The S-shaped $H_{\rm c2}$, that is the enhancement of $H_{\rm c2}$ occurs in the situation when UGe$_2$ crosses from FM1 to FM2,
and the moment is discontinuously increased and the Fermi surface drastically changes.~\cite{Tat01_mag,Hag02,Ter02}
Thus the enhancement of $H_{\rm c2}$ is not only due to the field dependence of the effective mass, but also due to the change of the Fermi surface.
The case of UGe$_2$ is more complicated than those of UCoGe and URhGe, 
in which the enhancement of $H_{\rm c2}$ or RSC can be simply explained by the field dependence of the effective mass.
Of course, it is important to clarify whether or not the Fermi surface and the renormalized band mass are invariant across $H_{\rm R}$ or $T_{\rm Curie}$ 
with microscopic experimental probes, such as de Haas-van Alphen effect.
This is left for future studies.

Resistivity measurements with fine tuning of field angle in high quality single crystals UCoGe lead us to find
new features with similarities and differences between two other known ferromagnetic superconductors URhGe and UGe$_2$.
They confirm that the magnetic field induces the FM instability accompanied by the enhanced $m^\ast$ of the conduction electrons,
consequently SC is stabilized, as illustrated in the S-shaped $H_{\rm c2}^b$ curve for $H\parallel b\mbox{-axis}$ in UCoGe.

We thank H. Harima, J. P. Brison, D. Braithwaite, V. P. Mineev and A. de Visser for helpful discussion.
This work was financially supported by French ANR project ECCE and CORMAT.


\end{document}